\documentclass[a4paper,twosides]{article}
\usepackage{multicol,multirow,graphics,fancyhdr,epstopdf}
\usepackage{amsmath,amsfonts,amssymb,bm,upgreek,mathrsfs,mathcomp}
\usepackage[pagewise,switch,columnwise]{lineno}
\usepackage[compress,nospace]{cite}
\usepackage[dvipsnames]{xcolor}
\usepackage{subfigure}%
\usepackage{graphicx}
\usepackage{booktabs}
\usepackage{amsmath}
\usepackage{CPL-2022}

\newcommand{\cplyear}{x} \newcommand{\cplvol}{xx}
\newcommand{\cplno}{x} \newcommand{\cplpagenumber}{xxxxxx}
 \newcommand{\cplpage}{\cplpagenumber-\thepage}

\begin{document}

\vspace* {-4mm} \begin{center}
\large\bf{\boldmath{QSEA: Quantum Self-supervised Learning with Entanglement Augmentation}}
\footnotetext{\hspace*{-5.4mm}$^{*}$Corresponding authors. Email: gaof@bupt.edu.cn
}
\\[5mm]
\normalsize \rm{}LingXiao Li$^{1,2}$, XiaoHui Ni$^{1,2}$, Jing Li$^{2}$, SuJuan Qin$^{1,2,3}$, and Fei Gao$^{1,2,3*}$
\\[3mm]\small\sl $^{1}$State Key Laboratory of Networking and Switching Technology, \\Beijing University of Posts and Telecommunications, Beijing, 100876, China

$^{2}$School of Cyberspace Security, Beijing University of Posts and Telecommunications, Beijing, 100876, China

$^{3}$National Engineering Research Center of Disaster Backup and Recovery, Beijing University of Posts and Telecommunications
\\[4mm]\normalsize\rm{}(Received xxx; accepted manuscript online xxx)
\end{center}

\vskip 1.5mm

\small{\narrower As an unsupervised feature representation paradigm, Self-Supervised Learning (SSL) uses the intrinsic structure of data to extract meaningful features without relying on manual annotation. Despite the success of SSL, there are still problems, such as limited model capacity or insufficient representation ability. Quantum SSL has become a promising alternative because it can exploit quantum states to enhance expression ability and learning efficiency. This letter proposes a Quantum SSL with entanglement augmentation method (QSEA). Different from existing Quantum SSLs, QSEA introduces an entanglement-based sample generation scheme and a fidelity-driven quantum loss function. Specifically, QSEA constructs augmented samples by entangling an auxiliary qubit with the raw state and applying parameterized unitary transformations. The loss function is defined using quantum fidelity, quantifying similarity between quantum representations and effectively capturing sample relations. Experimental results show that QSEA outperforms existing quantum self-supervised methods on multiple benchmarks and shows stronger stability in decorrelation noise environments. This framework lays the theoretical and practical foundation for quantum learning systems and advances the development of quantum machine learning in SSL.

\medskip
\noindent{\narrower{Keywords: Self-supervised learning, Quantum neural networks, Quantum machine learning, Data augmentation, Fidelity}}

\par}\vskip 3mm
\normalsize\noindent{\narrower{PACS: 03.67.Ac}}\\
\noindent{\narrower{DOI: \href{http://dx.doi.org/XX.XXXX/XXXX-XXXX/\cplvol/\cplno/\cplpagenumber}{XX.XXXX/XXXX-XXXX/\cplvol/\cplno/\cplpagenumber}}

\par}\vskip 5mm

\textit{Introduction.} Self-supervised learning (SSL) enables models to learn meaningful representations from unlabeled data by generating implicit supervision signals from raw inputs \cite{ssl1,ssl2,ssl3}. It has shown great success in fields like computer vision and natural language processing, where collecting large-scale labeled datasets is often expensive and labor-intensive. By reducing or eliminating the need for manual annotation, SSL offers a more efficient and cost-effective alternative to traditional supervised learning. 
\begin{figure}[htbp]
	\centering
	\setlength{\abovecaptionskip}{0.cm}
	\subfigure{
		\includegraphics[width=0.85\textwidth]{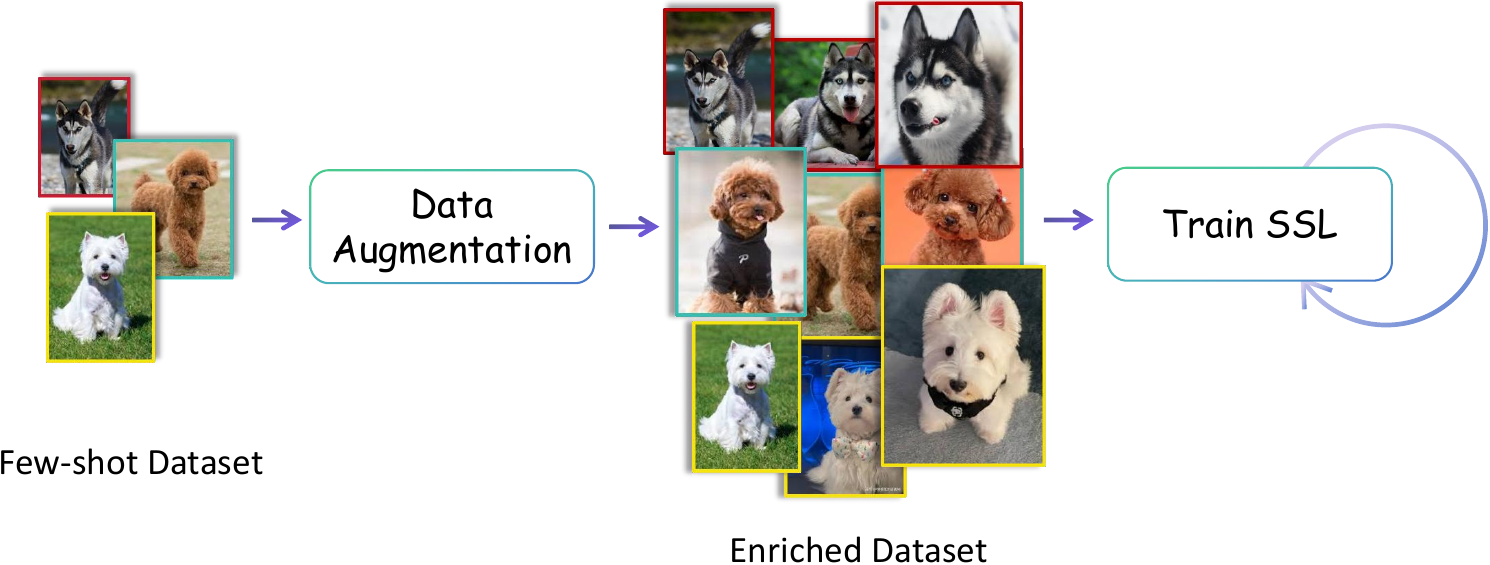}}
    \caption{The overall structure of SSL in few-shot datasets.}
\label{SSL}
\end{figure}
\medskip
However, in few-shot learning scenarios, where only limited samples are available, the performance of SSL models becomes heavily dependent on the quality and diversity of augmented data, the structure shown in FIG ~\ref{SSL}. Classical augmentation mechanism—such as cropping, color distortion, and geometric transformations \cite{vssl,gui,jing,simclr} —often fail to capture sufficient semantic variation, leading to overfitting and poor generalization. This limitation becomes especially critical when training data is sparse or lacks structure \cite{14,15,pu1,pu2}, which reduces the effectiveness of contrastive learning objectives commonly used in SSL.

In response to these challenges, emerging quantum machine learning algorithms are being explored as potential solutions. Quantum computing has gained significant attention due to its unique advantages, and the era of Noisy Intermediate-Scale Quantum (NISQ) \cite{xue,preskill} has arrived. With quantum resources offering unprecedented potential for solving complex tasks \cite{Arute,Huang1,Huang2}, quantum machine learning has emerged as a transformative research frontier in image processing, exemplified by pioneering models like Quantum Neural Networks\cite{li,cong,abb,beer,MLI}.  evolved to address the growing demand for label-efficient feature learning in quantum-enhanced systems.

Based on these advantages, quantum SSL has developed rapidly recently. Among these, Jaderberg \cite{jad} using amplitude encoding and contrastive learning to explore the expressive power of quantum models. Silver \cite{sil} introduces a quantum similarity detection network leveraging entangled input pairs to achieve stable learning under noise.  Yu \cite{yu} Designs a quantum unsupervised similarity learning framework with triplet construction and structure-aware circuit search for robust performance. To further explore the potential advantages of quantum computing, Kottahachchi \cite{kot} builds a fully quantum supervised contrastive learning model with quantum encoding, augmentation, and classification for few-shot image tasks. 


In this letter, we propose a Quantum Self-supervised learning with entanglement augmentation method (QSEA), a unified framework that leverages quantum entanglement and fidelity to enable effective self-supervised learning. QSEA constructs an enhanced dataset through quantum augmentation and optimizes representations via a fidelity-based loss, enabling robust model training under few-shot conditions. Specifically, QSEA introduces two key components: (1) Entanglement-based Augmentation (EA), where an auxiliary qubit is entangled with the raw state, followed by a parameterized unitary operation to generate diverse, informative augmented samples; (2) Fidelity-Driven Quantum Loss, which relies solely on quantum fidelity \cite{ban,bas,ram} to guide optimization. Maximizing similarity between raw and augmented states while minimizing alignment with unrelated samples. This design enables QSEA to learn semantically meaningful representations without requiring any classical labels. 
\begin{figure}[htbp]
	\centering
	\setlength{\abovecaptionskip}{0.cm}
	\subfigure{
		\includegraphics[width=0.85\textwidth]{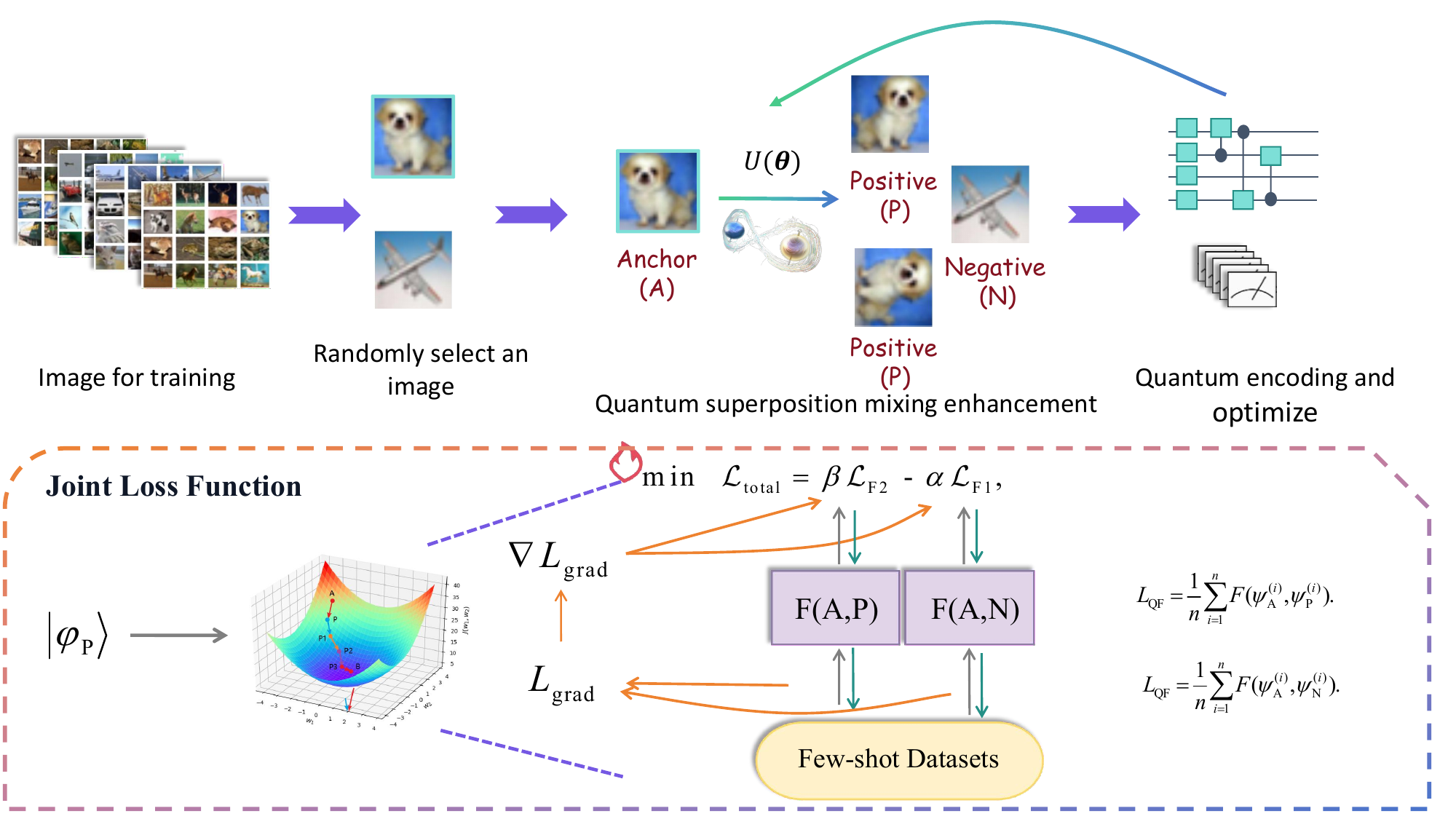}}
    \caption{The overall architecture of QSEA. The figure shows that for the training set, raw graphs (abbreviated as A) are randomly selected and data is enhanced through entanglement-based augmentation (EA). Enhanced samples are generated by introducing quantum random states and using parameter-containing unitary circuits $U(\boldsymbol{\theta})$ to introduce perturbations. Here, we regard all generated enhanced samples as positive samples  (abbreviated as P) and the rest as negative samples  (abbreviated as N). Finally, feature vectors are extracted through quantum circuit measurement and optimized using a quantum loss function, i.e. quantum fidelity contrast loss, to achieve training and testing applications of SSL.}
\label{0}
\end{figure}
Experimental results show that QSEA achieves high classification accuracy and surpasses other QSSL baselines on almost all standard benchmarks. It is worth noting that QSEA shows excellent robustness under the condition of simulating real device noise, with the accuracy only decreasing by 1.5\%, while the performance of \cite{yu} decreases by 6.5\% due to its inability to adapt to quantum perturbations. These results not only verify the advantages of superposition-based enhancement techniques in information fidelity but also highlight the unique role of entanglement modeling in optimizing feature representation. In summary, our research results lay a theoretical and practical foundation for quantum-classical hybrid learning architectures.

\medskip
\textit{Framework of quantum algorithm.} QSEA is a quantum self-supervised learning framework that leverages entanglement-based augmentation and fidelity-driven optimization to enable robust representation learning in few-shot scenarios. As illustrated in Figure~\ref{0}, the architecture is composed of three interconnected modules: (i) a classical-quantum hybrid encoder for transforming raw images into quantum representations; (ii) an entanglement-based augmentation mechanism that leverages quantum entanglement with an auxiliary qubit to generate enhanced samples; (iii) a quantum loss function solely based on fidelity, enabling self-supervised learning in the quantum domain.

Here, we use a classical-quantum hybrid encoder to encode images into quantum states, which is shown in Supplementary Materials Figure~\ref{bianmaAN}. The raw image is first normalized to the interval $[0,1]$ and processed by a lightweight classical encoder (e.g., ResNet-18), which extracts a compact feature vector of the raw image
\begin{equation}
	x_{\text{norm}} = \frac{x_{\text{A}}}{255},
\end{equation}
where, $x_{\text{norm}}$ represents the normalized pixel value, and $x_{\text{A}}$ is the raw pixel intensity. After normalization, the data is processed by a classical feature encoder. We use a lightweight ResNet-18 model \cite{kh,peng} as the encoder, which efficiently captures high-dimensional features while maintaining low computational cost. The raw images are compressed into a low-dimensional feature $v_i^{({\rm{A}})}$ through a classical encoder. The output feature vector is reduced to a lower-dimensional space, preparing the data for quantum amplitude encoding \cite{ae} and mitigating the curse of dimensionality. Then to transform the classical data into quantum states, for an $n$-qubit system, the Hilbert space dimension is $2^n$, and the classical data is encoded into a quantum amplitude state:
\begin{equation}
	|{\psi _{{\rm{A}}}}\rangle  = \sum\limits_{i = 0}^{{2^n} - 1} {v_i^{(A)}} |i\rangle,
\end{equation}
where, $\sum\limits_i | v_i^{(A)}{|^2} = 1.$  

\begin{figure*}[htbp]
	\centering
	\setlength{\abovecaptionskip}{0.cm}
	\subfigure{
		\includegraphics[width=0.75\textwidth]{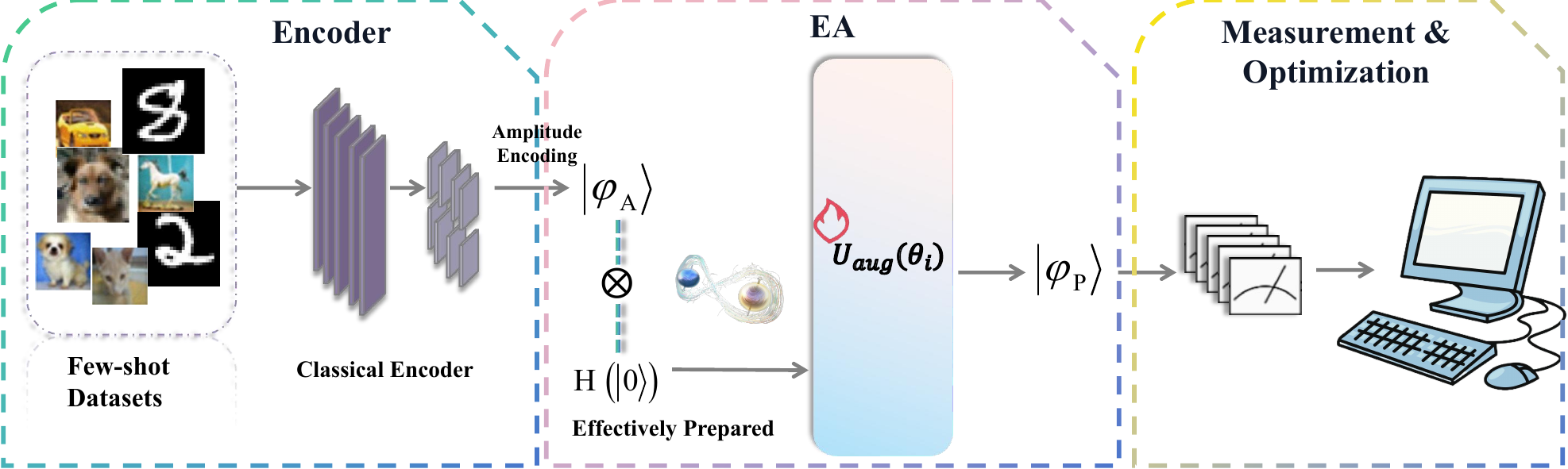}}
    \caption{The overall architecture and specific components of the Encoder and EA. QSEA uses a classical encoder to map classical data to quantum states through amplitude encoding to obtain the raw state $|{\psi _{{\rm{A}}}}\rangle $. An auxiliary qubit is entangled with the raw state $|{\psi _{{\rm{A}}}}\rangle$ via a unitary circuit ${U_{{\rm{aug}}}}(\boldsymbol{\theta} )$ to generate an enhanced state $|{\psi _{{\rm{P}}}}\rangle $. Finally, the feature vector is extracted through quantum circuit measurement, and the whole model training is performed. The degree of entanglement of the quantum random state during training is determined by the optimizable parameter $\boldsymbol{\theta}$.}
\label{01}
\end{figure*}

In this study, we introduce EA, a novel data augmentation approach leveraging quantum Entanglement.  The construct of EA can be seen in Figure~\ref{01}. By utilizing the potential advantages of quantum, EA aims to overcome the limitations of classical augmentation techniques, such as discrete operations and information loss, by exploiting the continuous parameter space of quantum states \cite{abd,per} to implement enhancements and increase information entropy \cite{ha,ts}. Unlike classical augmentation methods like random rotations or cropping \cite{ta1,ta2,zhong}, EA generates augmented states by entangling the raw state with an auxiliary qubit and applying a parameterized unitary circuit to introduce learnable perturbations.

\begin{figure*}[htbp]
	\centering
	\setlength{\abovecaptionskip}{0.cm}
	\subfigure{
		\includegraphics[width=0.7\textwidth]{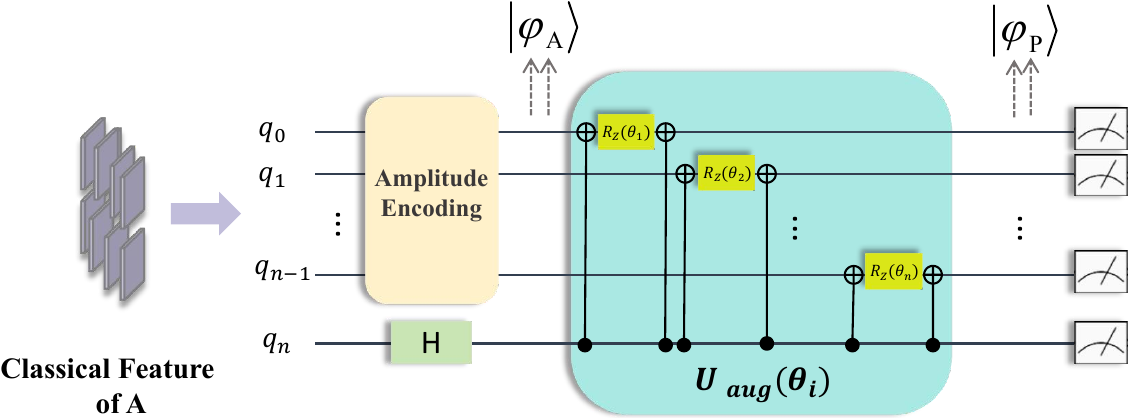}}
    \caption{EA process and circuit details. By introducing an input auxiliary bit, the raw state is entangled and perturbed. The auxiliary bit input is $|0\rangle $, and after the H gate operation, it is input into the unitary circuit ${U_{{\rm{aug}}}}(\boldsymbol{\theta} )$ together with the raw state to generate the enhanced state $|{\psi _{{\rm{P}}}}\rangle $, which is the positive sample.}
\label{zengq}
\end{figure*}

Here, an auxiliary qubit with an initial state of $|0\rangle$ first passes through the $H$ gate and is entangled with the raw state $|{\psi _{{\rm{A}}}}\rangle$ via a set of controlled quantum gates. A parameterized unitary operation ${U_{{\rm{aug}}}}(\boldsymbol{\theta} )$, composed of multi-qubit entangling layers and local rotations to generate an enhanced state $|{\psi _{{\rm{P}}}}\rangle $.
\begin{equation}
    \left|\psi_{\text {P}}\right\rangle=U_{\text {aug }}(\boldsymbol{\theta}) \cdot\left(\left|\psi_{\text {A}}\right\rangle \otimes|0\rangle\right),
\end{equation}
where $\boldsymbol{\theta}  \in [0,\pi /2]$ is learnable parameters. The mixing between the two states is governed by the learnable parameter $\theta$, which is updated during training. The circuit of EA diagram can be seen in Figure~\ref{zengq}. In summary, EA introduces a quantum-based approach to data augmentation. By using the continuous quantum parameter space, EA aims to enhance the diversity of samples and allows the model to be enhanced from multiple angles during the enhanced training process without losing the raw features. 



In quantum machine learning, the loss function is fundamental for guiding the model towards effective feature learning. Here, we propose a purely fidelity-based loss. In this letter, we propose a fidelity-based loss function that directly leverages quantum fidelity to quantify the similarity between quantum states. Specifically, the fidelity $F$ between two quantum states $\left| \psi  \right\rangle$ and $\left| \phi  \right\rangle $ is defined as the square of their inner product: $F = {\left| {\left\langle {\psi |\phi } \right\rangle } \right|^{\rm{2}}}.$  This measure captures how close two quantum states are, with $F=1$ means that $\left| \psi  \right\rangle$ and $\left| \phi  \right\rangle $ are the same sates and $F=0$ means that $\left| \psi  \right\rangle$ and $\left| \phi  \right\rangle $ are the orthogonal states. Here we divide the fidelity loss into two parts: $\mathcal{L}_{\text{F1}}$ for data enhancement optimization and $\mathcal{L}_{\text{F2}}$ for distance optimization of raw states and positive and negative states.

In order to improve the effectiveness of data enhancement features, quantum fidelity contrast loss ($\mathcal{L}_{\text{F1}}$) aims to make the enhanced quantum state closer to the raw state. This ensures that the representation learned by the model can retain the basic characteristics of the data while also introducing effective enhancements. The quantum fidelity $\mathcal{L}_{\text{F1}}$ loss is
\begin{equation}
    \max \quad L_{\mathrm{F} 1}=\frac{1}{N} \sum_{i=1}^{N} \frac{1}{1-F\left(\psi_{\mathrm{A}}^{(i)}, \psi_{\mathrm{P}}^{(i)}\right)},
\end{equation}
where $F(\psi _{{\rm{A}}}^{},\psi _{{\rm{P}}}^{})$ denotes the fidelity between raw states $\left| \psi _{{\rm{A}}}  \right\rangle $ and enhanced states$\left| \psi _{{\rm{P}}}  \right\rangle $. We hope that the raw state and the enhanced state are as similar as possible, so the optimization goal should be to make $F(\psi _{{\rm{A}}}^{},\psi _{{\rm{P}}}^{})$ as large as possible. However, this phenomenon will lead to the failure of data enhancement. This loss enforces a strong correlation between augmented samples and the raw data while minimizing the correlation with random, non-meaningful samples. Therefore, we choose to place the fidelity in the denominator of the loss function to prevent this from happening.

The $\mathcal{L}_{\text{F2}}$ for distance optimization of raw states and positive and negative states. For each training triplet—raw, positive, and negative—the loss function is designed to maximize the fidelity between the raw and its positive sample while minimizing the fidelity with a negative sample:
\begin{equation}
    \min \quad L_{\mathrm{F} 2}=\frac{1}{N} \sum_{i=1}^{N} F\left(\psi_{\mathrm{A}}^{(i)}, \psi_{\mathrm{N}}^{(i)}\right),
\end{equation}
where $F(\psi _{{\rm{A}}},\psi _{{\rm{N}}})$ denotes the fidelity between raw states $\psi _{{\rm{A}}}^{}$ and negtive states $\psi _{{\rm{N}}}^{}$. We hope that the raw state and the negative state are as inconsistent as possible, thus, the optimization goal of $\mathcal{L}_{\text{F2}}$ is to be as small as possible.

The total loss function is a weighted sum of the individual losses is:
\begin{equation}
    \min \quad \mathcal{L}_{\text {total }}=\beta \mathcal{L}_{\mathrm{F} 2}-\alpha \mathcal{L}_{\mathrm{F} 1},
\end{equation}
where $\alpha$ and $\beta$ are hyperparameters. This loss function ensures that the model optimizes quantum fidelity, captures higher-order entanglement correlations, making it suitable for quantum learning tasks that require both accuracy and resilience. For quantum circuits, we can obtain the fidelity calculation between two states by constructing a measurement circuit. The specific inner product measurement circuit refers to the construction of VQSD\cite{vqsd}. The specific fidelity calculation circuit can be seen in Supplementary Materials Figure~\ref{celiangF}.

In summary, we propose QSEA framework that integrates classical deep learning with quantum computing. By leveraging quantum data augmentation and a quantum loss function, our method achieves robust feature representations and improves generalization in noisy quantum environments.

\medskip
\textit{Numerical simulations.} By comparing with classical SSL algorithms and QSSL algorithms (Q-SupCon \cite{kot}, QSSL1 \cite{jad}, SLIQ \cite{sil}, QUSL \cite{yu}), the performance of QSEA based on quantum fidelity, entanglement entropy, and noise robustness in image classification tasks is comprehensively evaluated. The experiment uses three public datasets, MNIST, FMNIST, and CIFAR\_10, to demonstrate the advantages of QSEA in a few-shot learning environment, especially in improving model accuracy and robustness. The training set of each dataset contains only a small number of samples, simulating the challenge of learning effective features with limited samples in the case of data scarcity, and testing the generalization ability and robustness of the model.  Classification accuracy (Acc, Acc $= \frac{{{N_{{\rm{total}}}}}}{{{N_{{\rm{correct}}}}}}$), is used as the main evaluation indicator, each type of experiment was repeated 50 times to eliminate errors. In addition, in order to further explore the performance of QSEA, we carried out ablation experiments under different parameters, including the number of Qubits, the number of samples in each class, and the number of class.

\begin{figure*}[htbp]
	\centering
	\setlength{\abovecaptionskip}{0.cm}
	\subfigure{
		\includegraphics[width=0.75\textwidth]{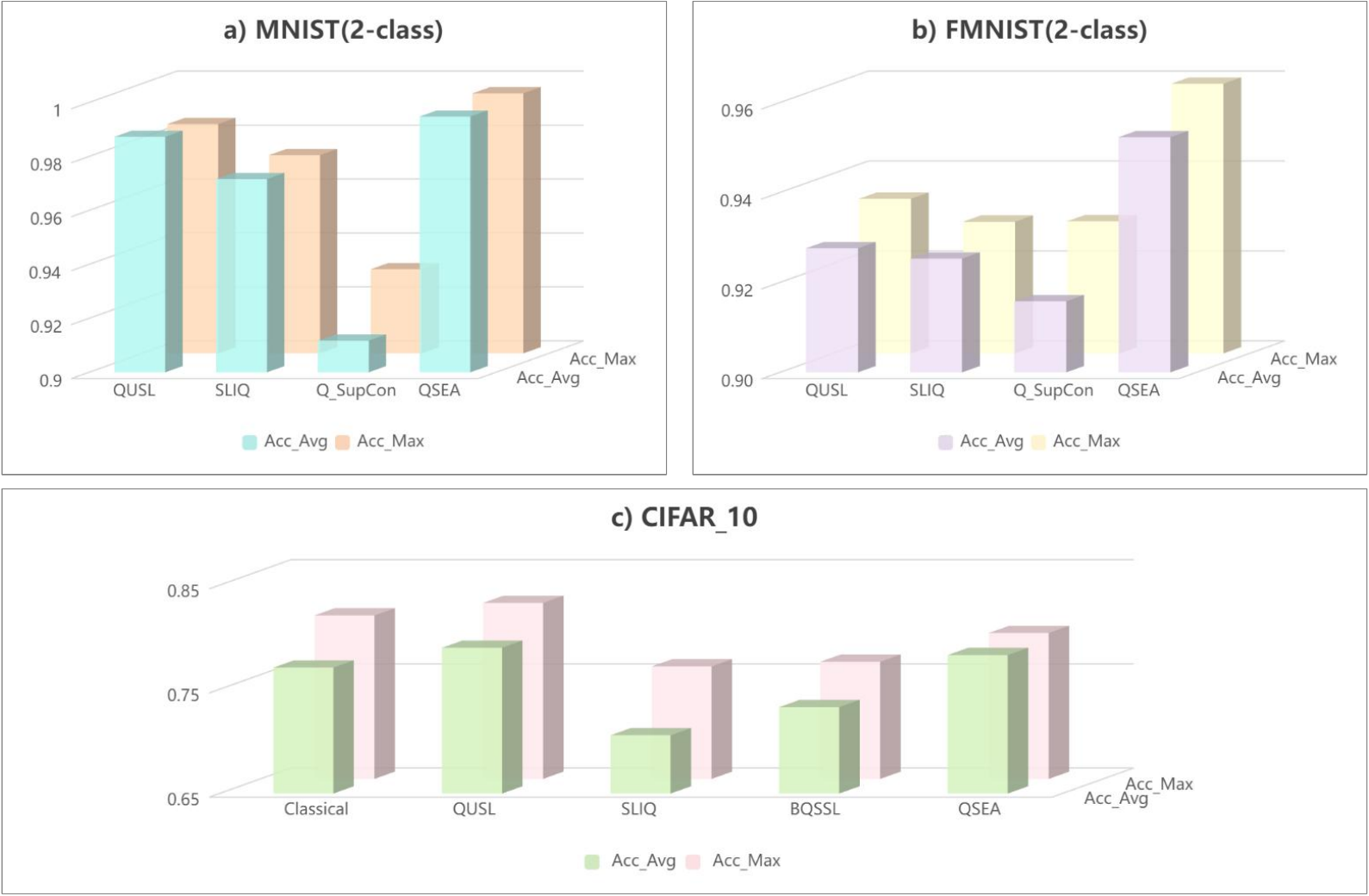}}
    \caption{The comparison test results include the maximum accuracy and average accuracy in 50 repeated experiments. Subfigure a) shows the accuracy of the quantum comparison method and the proposed method on the MNIST dataset; Subfigure b) shows the accuracy of the quantum comparison method and the proposed method on the FMNIST dataset; Subfigure c) shows the accuracy of the quantum comparison method and the proposed method on the CIFAR\_10 dataset.}
\label{1}
\end{figure*}

For comparative experiments, our experiments were conducted on three datasets, initially setting the number of Qubits to 8 and the number of samples per class to 50. QSEA performs well on almost all datasets, while other algorithms have mixed performance on different datasets, as shown in Figure~\ref{1}. Further highlighting the obvious advantage of QSEA in a few-shot learning environment.

In comparison, on the MNIST(2-class) and FMNIST(2-class) datasets, which can be seen in Figure~\ref{1} a) and b), QSEA once again demonstrated its excellent performance, with average accuracy of 99.76\% and 95.41\% respectively, far exceeding existing quantum self-supervised algorithms. On the MNIST dataset, although the QUSL algorithm showed good performance, it was still slightly inferior to the performance of QSEA. On the FMNIST dataset, QSEA's accuracy was as high as 96.35\%, about 4\% higher than multiple comparison methods. QSEA performs well on multiple datasets, while other algorithms have mixed performance on different datasets. On the CIFAR\_10 dataset, the classical baseline used a medium-scale fully connected neural network with 4 hidden layers and approximately 1 million parameters, employing the triplet strategy. As seen in Figure~\ref{1} c), QSEA achieved 79.01\% performance, outperforming other QSSL algorithms, second only to QUSL. The performance of SLIQ and QSSL1 is significantly inferior, further highlighting the obvious advantage of QSEA in a few-sample learning environment.



In addition to the few-shot learning, noise robustness is another critical evaluation metric in this experiment. Here, we choose to simulate the noise of NISQ equipment to simulate the performance of the algorithm in a noisy environment. We conducted simulated experiments using composite noise, including bit flip, phase flip, and depolarizing channel noise. As for the baseline, we choose the QUSL \cite{yu} that considers noise for comparison, and the experimental results are shown in Table ~\ref{noise}. The results on the CIFAR\_10 dataset show that QSEA performs well in both noise-free and noisy environments, with a difference of approximately 1.5\%, demonstrating its exceptional noise robustness. This performance highlights the stability and reliability of QSEA in the face of various environmental interferences. In contrast, although QUSL performs well in the noise-free environment, its accuracy drops in noisy environments, indicating its vulnerability to noise interference.

\begin{table}[htbp]
  \centering
  \caption{Comparison results with and without noise}
    \begin{tabular}{c|ccccc}
    \toprule[1pt]
          & Model & Acc\_Max & Acc\_Avg & $\Delta$Acc\_Max & $\Delta$Acc\_Avg \\
    \midrule
    \multirow{2}{*}{QUSL \cite{yu}} & Noise\_Free & 80.21\% & 78.78\% & \multirow{2}{*}{4.41\%} & \multirow{2}{*}{6.52\%} \\
          & Noise & 75.80\% & 72.26\% &       &  \\
    \midrule
    \multirow{2}{*}{QSEA (Ours)} & Noise\_Free & 78.62\% & 78.21\% & \multirow{2}{*}{\textbf{1.50\%}} & \multirow{2}{*}{\textbf{1.37\%}} \\
          & Noise & 77.12\% & 76.84\% &       &  \\
    \bottomrule[1pt]
    \end{tabular}%
  \label{noise}%
\end{table}%

\begin{figure*}[htbp]
	\centering
	\setlength{\abovecaptionskip}{0.cm}
	\subfigure{
		\includegraphics[width=0.99\textwidth]{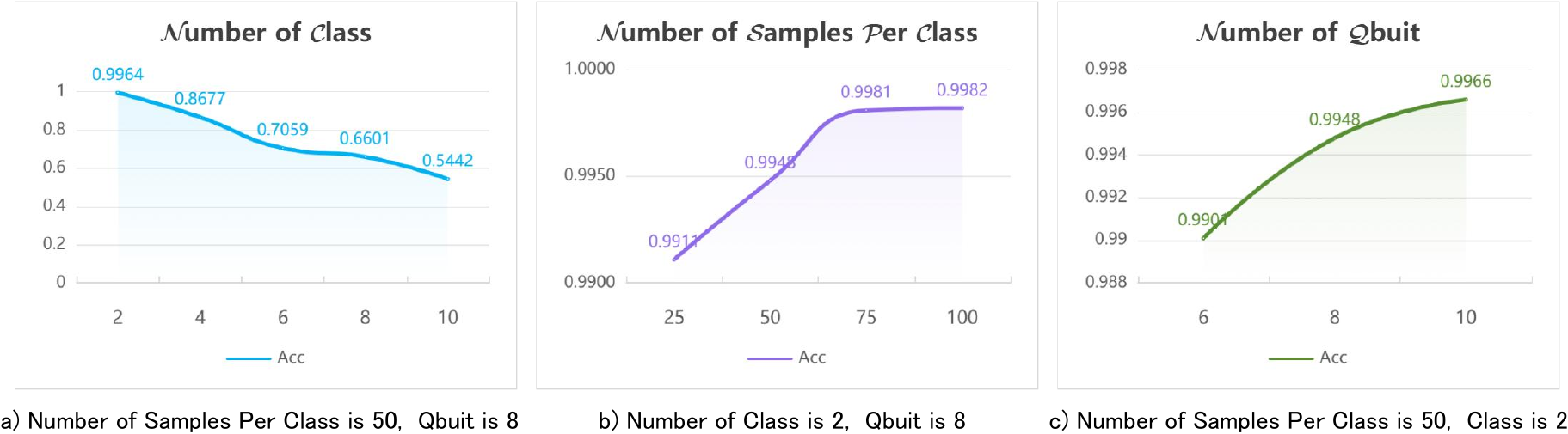}}
    \caption{Acc analysis of QSEA model. Subfigure a) shows that the classification accuracy of QSEA decreases significantly as the number of categories increases, highlighting the model's sensitivity to category expansion; Subfigure b) shows the robustness of the QSEA model, which can maintain a high classification accuracy even when the number of samples decreases; Subfigure c) shows that the performance of QSEA eventually stabilizes as the number of qubits increases, demonstrating the ability of QSEA to model high-dimensional features. }
\label{2}
\end{figure*}

For ablation experiments, we explored the impact of these factors on the performance of the QSEA model by changing the number of categories, the number of qubits, and the number of samples per category. The results show that these factors affect the performance of the model to varying degrees, as shown in Figure~\ref{2}.

As the number of classes increases, the classification accuracy of the model gradually decreases. This can be attributed to the increasing complexity of the classification task. With more classes, the model is required to differentiate between more categories, making the classification task more challenging. Particularly in situations with limited samples, the model may struggle to learn distinct features for each class, resulting in a drop in accuracy. Nevertheless, even with a larger number of classes, QSEA still maintains a relatively high accuracy, indicating that the model can effectively capture key information across multiple classes. In the experiment with the number of qubits, we observed a positive impact on the model's performance as the number of qubits increased. 

As the number of qubits increased from 6 to 10, the accuracy of the model steadily improved. This result suggests that increasing the number of qubits provides the model with more computational power and a stronger capacity for feature representation, allowing it to better handle the complex patterns within the data. The increase in qubits essentially expands the latent space of the model, leading to higher accuracy. The experiment on the number of samples per class showed the most significant impact on the performance. 

As the number of samples per class increased, the accuracy of the model improved steadily. This indicates that more training samples allow the model to gather richer information, thus improving its ability to learn the underlying patterns in the data. An increase in sample size helps to reduce overfitting during training and provides more data for better generalization, ultimately leading to more accurate predictions when faced with unseen samples. 
Through comprehensive experimental analysis, QSEA has shown significant advantages over classical and quantum SSL algorithms on a variety of datasets and still maintains high accuracy and robustness when dealing with different numbers of categories, numbers of quantum bits, and numbers of samples, fully verifying its strong application potential in complex tasks.

\textit{Conclusion and outlook.} This study proposes the QSEA framework to systematically solve the core challenges of insufficient feature diversity and limited noise robustness of traditional SSL in few-shot scenarios. QSEA provides a new paradigm for QSSL through the innovative combination of quantum superposition enhancement and quantum loss function design. Experimental results show that QSEA exhibits significant stability advantages in a few-shot learning. The core lies in the continuous parameterization space of EA and the explicit high-order correlation modeling of entanglement entropy regularization loss, which effectively alleviates the risk of overfitting under few-shot. The quantum loss function constrains positive and negative sample pairs through fidelity comparison. The two mechanisms together to make QSEA perform well in multiple scenarios. Future work will focus on exploring the optimization of the hardware compatibility of QSEA on real NISQ devices. For example, further reduction of the circuit depth through quantum gate decomposition and dynamic subsystem partitioning strategies is needed to cope with the noise constraints and computing resource limitations of quantum hardware, thereby promoting the implementation of the quantum enhancement learning paradigm into practical engineering applications.

\medskip
\textit{Acknowledgements.} This work is supported by National Natural Science Foundation of China (Grant Nos. 62371069, 62372048, 62272056), and BUPT Excellent Ph.D. Students Foundation(CX20241055). The numerical simulations are mainly performed by using MindSpore Quantum \cite{mq} 0.9.11 framework. The updated code is avilable at \cite{git}.

\newpage

\vspace{5ex}
\section*{A. Supplementary Materials}
\renewcommand{\thefigure}{\Alph{figure}}
\setcounter{figure}{0} 
\begin{figure*}[htbp]
	\centering
	\setlength{\abovecaptionskip}{0.cm}
	\subfigure{
		\includegraphics[width=0.99\textwidth]{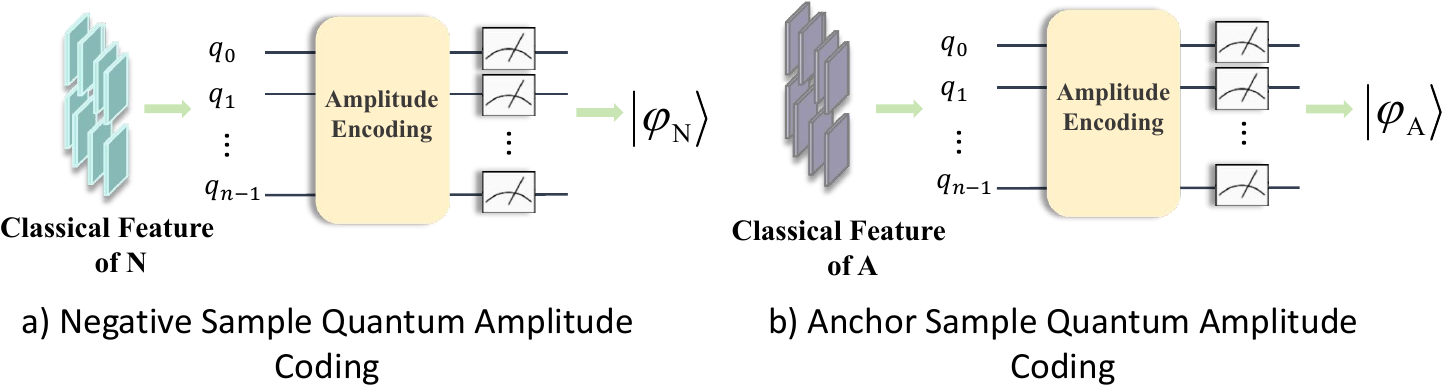}}
    \caption{The details of the quantum encoder. }
\label{bianmaAN}
\end{figure*}

\begin{figure*}[htbp]
	\centering
	\setlength{\abovecaptionskip}{0.cm}
	\subfigure{
		\includegraphics[width=0.99\textwidth]{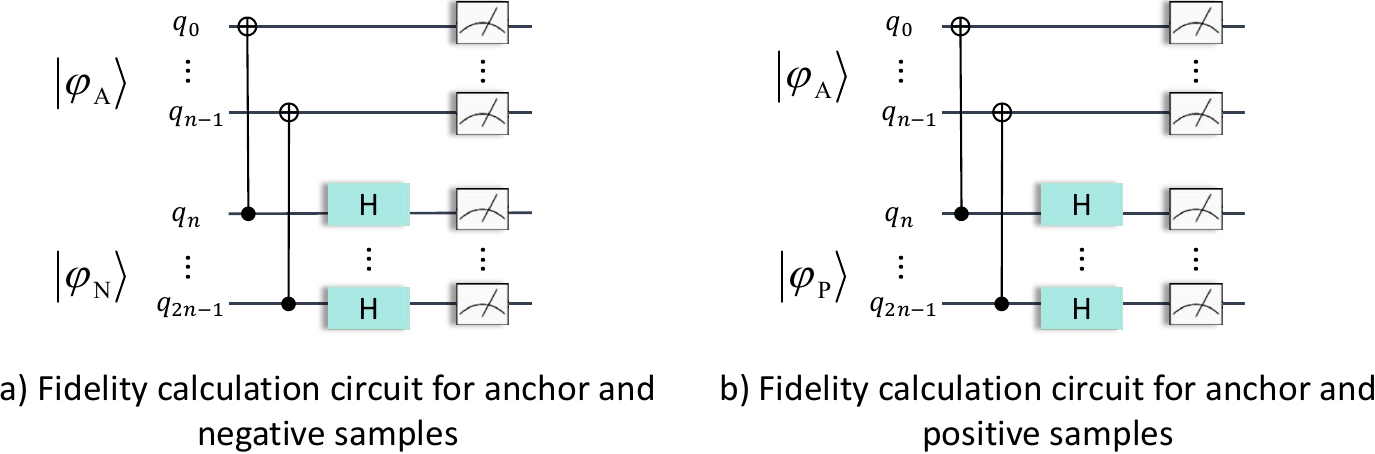}}
    \caption{Fidelity measurement circuit. Fidelity measurement lines between raw states and positive states, and between raw states and negative states.}
\label{celiangF}
\end{figure*}


\begin{thebibliography}{99}\footnotesize
\itemsep=-1pt plus.2pt minus.2pt

\bibitem{ssl1} Liu X, Zhang F, Hou Z, et al., 2021, \textit{IEEE Trans. Knowl. Data Eng.} \textbf{35}(1) 857--876

\bibitem{ssl2} Zhai X, Oliver A, Kolesnikov A, et al., 2019, \textit{Proc. IEEE/CVF Int. Conf. Comput. Vis.} 1476--1485

\bibitem{ssl3} Gui J, Chen T, Zhang J, et al., 2024, \textit{IEEE Trans. Pattern Anal. Mach. Intell.}

\bibitem{vssl} Kolesnikov A, Zhai X and Beyer L, 2019, \textit{Proc. IEEE/CVF Conf. Comput. Vis. Pattern Recognit.} 1920--1929

\bibitem{gui} Gui J, Chen T, Zhang J, et al., 2024, \textit{IEEE Trans. Pattern Anal. Mach. Intell.}

\bibitem{jing} Jing L and Tian Y, 2020, \textit{IEEE Trans. Pattern Anal. Mach. Intell.} \textbf{43}(11) 4037--4058

\bibitem{simclr} Chen T, Kornblith S, Norouzi M, et al., 2020, \textit{Int. Conf. Mach. Learn., PMLR} 1597--1607

\bibitem{18} Wang G, Wang K, Wang G, et al., 2021, \textit{Proc. IEEE/CVF Int. Conf. Comput. Vis.} 9505--9515

\bibitem{14} Henaff O, 2020, \textit{Int. Conf. Mach. Learn., PMLR} 4182--4192

\bibitem{15} Oord A, Li Y and Vinyals O, 2018, \textit{arXiv preprint arXiv:1807.03748}

\bibitem{pu1} Wang Y, Song R, Li L, et al., 2025, \textit{Appl. Soft Comput.} 113101

\bibitem{pu2} Wang Y, Li L, Tang Y, et al., 2025, \textit{IEEE Trans. Intell. Transp. Syst.}


\bibitem{19} He K, Chen X, Xie S, et al., 2022, \textit{Proc. IEEE/CVF Conf. Comput. Vis. Pattern Recognit.} 16000--16009

\bibitem{23} Havlíček V, Córcoles A D, Temme K, et al., 2019, \textit{Nature} \textbf{567}(7747) 209--212

\bibitem{xue} Xue C, Chen Z Y, Wu Y C, et al., 2021, \textit{Chinese Phys. Lett.} \textbf{38}(3) 030302

\bibitem{preskill} Preskill J, 2018, \textit{Quantum} \textbf{2} 79

\bibitem{Arute} Arute F, Arya K, Babbush R, et al., 2019, \textit{Nature} \textbf{574}(7779) 505--510

\bibitem{Huang1} Huang H Y, Broughton M, Cotler J, et al., 2022, \textit{Science} \textbf{376}(6598) 1182--1186

\bibitem{Huang2} Huang H Y, Broughton M, Mohseni M, et al., 2021, \textit{Nat. Commun.} \textbf{12}(1) 2631

\bibitem{li} Li L, Li J, Song Y, et al., 2025, \textit{Sci. China Phys. Mech. Astron.} \textbf{68}(1) 1--9

\bibitem{cong} Cong I, Choi S and Lukin M D, 2019, \textit{Nat. Phys.} \textbf{15}(12) 1273--1278

\bibitem{abb} Abbas A, Sutter D, Zoufal C, et al., 2021, \textit{Nat. Comput. Sci.} \textbf{1}(6) 403--409

\bibitem{beer} Beer K, Bondarenko D, Farrelly T, et al., 2020, \textit{Nat. Commun.} \textbf{11}(1) 808

\bibitem{MLI} Ni X H, Cai B B, Liu H L, Qin S J, Gao F and Wen Q Y 2024 \textit{Adv. Quantum Technol.} \textbf{7} 2300419

\bibitem{jad} Jaderberg B, Anderson L W, Xie W, et al., 2022, \textit{Quantum Sci. Technol.} \textbf{7}(3) 035005

\bibitem{sil} Silver D, Patel T, Ranjan A, et al., 2023, \textit{Proc. AAAI Conf. Artif. Intell.} \textbf{37}(8) 9846--9854

\bibitem{yu} Yu L H, Li X Y, Chen G, et al., 2024, \textit{Expert Syst. Appl.} \textbf{258} 125112

\bibitem{kot} Kottahachchi Kankanamge Don A and Khalil I, 2025, \textit{ACM Trans. Quantum Comput.} \textbf{6}(1) 1--24

\bibitem{ab} Abouelnaga Y, Ali O S, Rady H, et al., 2016, \textit{Proc. 2016 Int. Conf. Comput. Sci. Comput. Intell. (CSCI), IEEE} 1192--1195

\bibitem{deng} Deng L, 2012, \textit{IEEE Signal Process. Mag.} \textbf{29}(6) 141--142

\bibitem{xiao} Xiao H, Rasul K and Vollgraf R, 2017, \textit{arXiv preprint arXiv:1708.07747}

\bibitem{ban} Banchi L, Braunstein S L and Pirandola S, 2015, \textit{Phys. Rev. Lett.} \textbf{115}(26) 260501

\bibitem{bas} Bason M G, Viteau M, Malossi N, et al., 2012, \textit{Nat. Phys.} \textbf{8}(2) 147--152

\bibitem{ram} Rams M M and Damski B, 2011, \textit{Phys. Rev. Lett.} \textbf{106}(5) 055701

\bibitem{deng} Deng D L, Li X and Das Sarma S, 2017, \textit{Phys. Rev. X} \textbf{7}(2) 021021

\bibitem{has} Hastings M B, González I, Kallin A B, et al., 2010, \textit{Phys. Rev. Lett.} \textbf{104}(15) 157201

\bibitem{kh} Khan R U, Zhang X, Kumar R, et al., 2018, \textit{Proc. 2018 Int. Conf. Comput. Artif. Intell.} 86--90

\bibitem{peng} Peng C, Tian T, Chen C, et al., 2021, \textit{Neural Netw.} \textbf{137} 188--199

\bibitem{ae} Davis J A, Cottrell D M, Campos J, et al.,1999, \textit{Appl. optics}  \textbf{38}(23) 5004-5013.

\bibitem{abd} Abdessaied N and Drechsler R, 2016, \textit{Optim. Complex. Anal.}, Springer, Cham

\bibitem{per} Peres A, 1985, \textit{Phys. Rev. A} \textbf{32}(6) 3266

\bibitem{ha} Harte J and Newman E A, 2014, \textit{Trends Ecol. Evol.} \textbf{29}(7) 384--389

\bibitem{ts} Tsai D Y, Lee Y and Matsuyama E, 2008, \textit{J. Digit. Imaging} \textbf{21} 338--347

\bibitem{ta1} Takahashi R, Matsubara T and Uehara K, 2018, \textit{Asian Conf. Mach. Learn., PMLR} 786--798

\bibitem{ta2} Takahashi R, Matsubara T and Uehara K, 2019, \textit{IEEE Trans. Circuits Syst. Video Technol.} \textbf{30}(9) 2917--2931

\bibitem{zhong} Zhong Z, Zheng L, Kang G, et al., 2020, \textit{Proc. AAAI Conf. Artif. Intell.} \textbf{34}(07) 13001--13008

\bibitem{vqsd} LaRose, Ryan, Tikku, Arkin,  O’Neel-Judyet al.,2019, \textit{npj Quantum Inf.} \textbf{5} 57

\bibitem{he} He Z, Rakin A S and Fan D, 2019, \textit{Proc. IEEE/CVF Conf. Comput. Vis. Pattern Recognit.} 588--597

\bibitem{ga} Garcia L P F, Lehmann J, de Carvalho A C, et al., 2019, \textit{Knowl.-Based Syst.} \textbf{163} 693--704

\bibitem{foi} Foi A, Trimeche M, Katkovnik V, et al., 2008, \textit{IEEE Trans. Image Process.} \textbf{17}(10) 1737--1754

\bibitem{mc} McHutchon A and Rasmussen C, 2011, \textit{Adv. Neural Inf. Process. Syst.} \textbf{24}

\bibitem{vl} Vlatakis-Gkaragkounis E V, Flokas L and Piliouras G, 2021, \textit{Adv. Neural Inf. Process. Syst.} \textbf{34} 2373--2386

\bibitem{no} Nouiehed M, Sanjabi M, Huang T, et al., 2019, \textit{Adv. Neural Inf. Process. Syst.} \textbf{32}


\bibitem{mq} Xu X, Cui J, Cui Z, et al. \textit{arXiv preprint arXiv:2406.17248}
\bibitem{git} https://gitee.com/lilingxiao2025/q-ssl

\end{thebibliography}
\end{document}